\documentclass{appolb}
\usepackage{amssymb,amsfonts,amsmath,amsthm,lineno}

\usepackage[showonlyrefs]{mathtools}
\mathtoolsset{showonlyrefs=true}

\MakeRobust{\eqref}

\usepackage{hyperref}

\newcommand{\mR}{\mathbb{R}}
\newcommand{\mZ}{\mathbb{Z}}
\newcommand{\mC}{\mathbb{C}}
\newcommand{\dsp}{\displaystyle}

\newcommand{\al}{\alpha}
\newcommand{\la}{\lambda}
\newcommand{\p}{\partial}
\newcommand{\w}{\omega}
\newcommand{\W}{\Omega}
\newcommand{\Ac}{\mathcal{A}}
\newcommand{\Bc}{\mathcal{B}}
\newcommand{\Hc}{\mathcal{H}}
\newcommand{\K}{\mathcal{K}}
\newcommand{\Lc}{\mathcal{L}}
\newcommand{\ov}{\overline}
\newcommand{\con}{\mathrm{const}}
\newcommand{\1}{\mathbf{1}}
\newcommand{\Lor}{\mathcal{L}^\uparrow_+}

\newcommand{\dA}{\dot{A}}
\newcommand{\dB}{\dot{B}}

\DeclareMathOperator{\Rp}{Re}
\DeclareMathOperator{\Ip}{Im}

\DeclareMathOperator{\id}{id}
\DeclareMathOperator{\Ker}{Ker}

\DeclareMathOperator{\sgn}{sgn}
\DeclareMathOperator{\arcosh}{arcosh}

\begin{document}

\title{Remarks on mathematical structure\\ of the Staruszkiewicz theory}
\author{Andrzej Herdegen\thanks{e-mail address:
herdegen@th.if.uj.edu.pl}
\address{Institute of Theoretical Physics\\
    Jagiellonian University\\
    ul.\,S.\,{\L}ojasiewicza 11,\
    30-348 Krak\'{o}w\\
    Poland}}

\maketitle

\begin{abstract}
 The purpose of this note is threefold: (i) to recall (with some points made more explicit) the mathematical Weyl algebra model formulation, given before,  of the Staruszkiewicz theory of quantum Coulomb field; (ii)~to add some new elements to the discussion of the representation of the Lorentz group within this model; (iii) to comment on some statements on the structure of the theory which appeared recently.
\end{abstract}

\section{Introduction}

In this note we want to comment on the mathematical structure of the  Staruszkiewicz theory of the long-range asymptotics of electromagnetic field. This theory was postulated by its author in Ref.\ \cite{sta89} in the form of a set of algebraic conditions on its basic operators acting in some Hilbert space. While having interesting physical motivation, that formulation could leave some doubts whether its axioms have a consistent mathematical status, as no closed mathematical model, known to exist, was shown to satisfy these axioms. However, the nature of the postulates indicated in a rather obvious way where one should look for such a model, namely in the range of Weyl algebras of canonical commutation relations, but based not on a symplectic vector space, but rather on a symplectic additive abelian group. The present author, being interested in related problems,\footnote{For the present author's approach to the infrared problems in quantum electrodynamics see a recent account in Ref.\ \cite{her17}, as well as Section 4 in Ref.\ \cite{her05}.} suggested a concrete solution  in Ref.\ \cite{her94}, and later in Ref.\ \cite{her05} it was shown that indeed Staruszkiewicz's axioms, reformulated in an appropriate way, are implemented in a specific representation of certain Weyl algebra. As this  mathematical model is fully consistent, no doubts remain.

However, as it seems, the message of the Weyl algebra has not been given consideration by some authors interested in the Staruszkiewicz theory, which is a pity both from the fundamental, as well as the practical point of view. This is exemplified by two recent articles, Refs.\ \cite{waw18} and \cite{waw22}, the former is fundamentally wrong, and the latter contains, apart from one result worth noting, largely redundant and misleading considerations. In our opinion these articles would not have come to being (at least in the present form), had their author taken the lesson of the Weyl algebra into account.

In view of the above, we find it desirable to recall the main structure points of the Weyl algebra model once more, some of them more explicitly than before. We concentrate here on mathematics of the model; for physical interpretation which makes it the physical theory put forward by Staruszkiewicz, we refer the reader to Section 3 of Ref.~\cite{her05}, and to original articles by Staruszkiewicz. The construction of the model is explained in Sections  \ref{alg} and \ref{repr}. In Section \ref{corresp} we describe, more exhaustively than in \cite{her05}, the correspondence of the model with Staruszkiewicz's axioms. Section~\ref{subspaces} is devoted to the description of the representation of the Lorentz group acting in the model. Besides known facts, it also contains some new contributions.  Then, in Section \ref{comments}, we comment on Ref.\ \cite{waw18} and \cite{waw22}. Some technical information and proofs are shifted to Appendix.

\section{Algebra}\label{alg}

Let $\Lc$ be a set of pairs $(D,c)$, where $D(l)$ and $c(l)$ are real  functions on the future light cone, smooth outside the vertex, $D$ are homogeneous of degree $0$, $c$ are homogeneous of degree $-2$, and restricted by the condition\footnote{Notation of the integral and a few elementary facts on homogeneous functions on the light cone are gathered in Appendix \ref{app_hom}.}
\begin{equation}\label{alg_nc}
 n_c\equiv\frac{1}{4\pi e}\int c(l)\,d^2l\in\mZ\,,
\end{equation}
where $e$ is a positive constant, which in the Staruszkiewicz theory acquires the meaning of the elementary charge (we use units $\hbar=1$, $c=1=$ speed of light). Further, let $\Lc_0$ denote the subset of $\Lc$ composed of elements $(D,c)$ with $n_c=0$. The set $\Lc$ has the structure of an abelian group,  with the addition defined by
\[
 (D_1,c_1)+(D_2,c_2)=(D_1+D_2,c_1+c_2)\,,
\]
and $\Lc_0$ is then a vector space, with obvious definition of multiplication by scalars. We define on $\Lc$ a symplectic mapping $\sigma:\Lc\times\Lc\mapsto \mR$ by
\begin{equation}
 \sigma(D_1,c_1;D_2,c_2)
 =\frac{1}{4\pi}\int \big[D_1(l)c_2(l)-D_2(l)c_1(l)\big]\,d^2l\,.
\end{equation}
This mapping is additive in each of its arguments, bilinear on $\Lc_0\times\Lc_0$, and is nondegenerate in the sense of the following equivalence:
\[
 \exp[i\sigma(D,c;D',c')]=1\quad \forall (D',c')\in\Lc
 \qquad \Longleftrightarrow\qquad (D,c)=(0,0)\,.
\]
It follows that the symplectic group $(\Lc,\sigma)$ defines a unique Weyl $C^*$-algebra generated by elements $W(D,c)$ satisfying the relations
\begin{equation}\label{alg_weyl}
\begin{gathered}
 W(D_1,c_1)W(D_2, c_2)
 =\exp\big[\tfrac{i}{2}\sigma(D_1, c_1; D_2, c_2)\big]
 W(D_1+D_2,c_1+c_2)\,,\\[1ex]
 W(D, c)^*=W(-D, -c)\,,\qquad W(0,0)=\1\,.
 \end{gathered}
\end{equation}
This algebra, which will be denoted $\Ac$, is simple.
These fundamental statements follow from Corollaries (4.23) and (4.24) of Ref.\ \cite{mstv73}, which generalize their versions most often used, when   $(\Lc,\sigma)$ is a nondegenerate symplectic vector space. The algebra is thus consistently defined for an \emph{a priori} arbitrary value of $e$.
Elements $W(D,c)$ with $(D,c)\in\Lc_0$ generate a $C^*$-subalgebra~$\Ac_0$.

There exists a simple characterization of $\Lc_0$ and $\Ac_0$, to become of  importance in what follows. For $n_c=0$ one can use equivalence \eqref{app_cF} in Appendix~\ref{app_hom}, to write $c=\p^2F$, where $F$ is homogeneous of degree $0$; this $F$ is unique up to the addition of constants. Therefore, elements of $\Lc_0$ may be written as $(D,\p^2F)$, and generators of $\Ac_0$ as $W(D,\p^2F)$. Denote by $H$ the future unit hyperboloid in Minkowski space. Then the whole algebra $\Ac$ is generated by elements $W(D,\p^2F)$ and $W(0,c_v)$, where
\begin{equation}\label{alg_cv}
 c_v(l)=\frac{e}{(v\cdot l)^2}\,,\quad v\in H\,.
\end{equation}
Note that $c_u-c_v=\p^2 F_{vu}$ (with $F_{vu}$ explicitly given in \eqref{subspaces_cc} below). Therefore, for generation of the algebra, next to the elements $W(D,\p^2 F)$, it is sufficient to use only one element $W(0,c_v)$.

The restricted Lorentz group $\Lor$ acts by a group of automorphisms on~$\Ac$. For any function $f$ on the future light cone and $\Lambda\in\Lor$ we denote
\begin{equation}\label{alg_T}
 [T_\Lambda f](l)=f(\Lambda^{-1}l)\,,
\end{equation}
and define the automorphism $\al_\Lambda$ by its action on the generating elements:
\begin{equation}
 \al_\Lambda(W(D, c))=W(T_\Lambda D, T_\Lambda c)\,,\quad \text{so}\quad \al_\Lambda\al_{\Lambda'}=\al_{\Lambda\Lambda'}\,.
\end{equation}
The subalgebra $\Ac_0$ is invariant under the action of these automorphisms.

\section{Lorentz-covariant representations}\label{repr}

Each $C^*$-algebra has Hilbert space representations by bounded operators.
To construct a cyclic, Lorentz-covariant representation of $\Ac$, it is sufficient to define a Lorentz-invariant state (i.e.\ a positive, normalized linear functional) on the algebra. By the GNS procedure this state then gives rise to a unique representation (see e.g.\ \cite{bro87}), in which Lorentz transformations are implemented by a unitary representation.

For that purpose, consider the space of complex, smooth functions on the future light cone, homogeneous of degree $0$. Next, identify (by an equivalence relation) functions differing by an additive constant; the elements will be denoted $[G]$ etc. On that space we define the positive definite scalar product\footnote{In Ref.\ \cite{her05} there is a misprint at this place, the conjugation sign over the left argument is missing.}
\begin{equation}
 ([G_1],[G_2])_\K=\frac{1}{4\pi}\int\ov{G_1(l)}\p^2 G_2(l)\,d^2l
 =-\frac{1}{4\pi}\int \ov{\p G_1(l)}\cdot\p G_2(l)\,d^2l\,;
\end{equation}
the differential and integral operations appearing above are explained in Appendix \ref{app_hom}. The completion of this space in the topology of the product is a Hilbert space, which we denote $\K$. For real, smooth, homogeneous of degree $0$ functions $D$ and $F$ we introduce notation
\begin{equation}\label{repr_j}
 j(D,F)=\kappa^{-\frac{1}{2}}D-i\kappa^{\frac{1}{2}}F\,,
\end{equation}
where $\kappa$ is any positive real number.
It is then not difficult to see that the following prescription defines a~Lorentz-invariant state $\w$ on the algebra $\Ac$:
\begin{equation}\label{repr_state}
\begin{aligned}
 \w\big(W(D, c)\big)&=0\qquad \text{for}\qquad n_c\neq 0\,,\\
 \w\big(W(D,\p^2 F)\big)
 &=\exp\big[-\tfrac{1}{4}\|[j(D,F)]\|^2_\K\big]\\
 &=\exp\big[-\tfrac{1}{4}\big(\kappa^{-1}\|[D]\|_\K^2+\kappa\,\|[F]\|_\K^2\big)\big]
 \end{aligned}
\end{equation}
---we sketch the proof of this fact in Appendix \ref{app_state}.
Ten the GNS procedure ensures that there exist a Hilbert space $\Hc$, a~vector $\W\in\Hc$ and a~representation $\Ac\ni A\mapsto \pi(A)\in\Bc(\Hc)$, such that $\W$ is cyclic for $\pi(\Ac)$, and for each $A\in \Ac$:
\begin{equation}\label{repr_gns}
 \w(A)=(\W,\pi(A)\W)\,.
\end{equation}
The representation $\pi$ is unique up to unitary equivalence. As $\Ac$ is simple, the representation is faithful.
It should be clear that all objects $j$, $\w$, $\Hc$, $\W$ and $\pi$ depend on the parameter $\kappa$, but not to burden notation we do not indicate this explicitly. It may be shown that representations with different $\kappa$'s are nonequivalent. Physical demands Staruszkiewicz imposes on his theory are equivalent to the choice
\begin{equation}\label{repr_kappa}
 \kappa=\frac{2}{\pi}\,,
\end{equation}
which should be set for Staruszkiewicz's theory in further formulas. We explain this choice in the next section.

While the GNS procedure guarantees mathematical consistency of the representation, it is desirable to describe it in more explicit terms. As the representation $\pi$ is fixed and faithful, which implies that the map $\Ac\mapsto \pi(\Ac)$ is a ${}^*$-isomorphism, to simplify notation we can omit its symbol and in what follows identify
\begin{equation}
 W(D,c)=\pi(W(D,c))\,.
\end{equation}
Let \mbox{$\Hc_n\subset\Hc$} denote the closure of the subspace spanned by all vectors of the form \mbox{$W(D, c)\W$} with $n_c=n$. Then $\Hc$ is the orthogonal direct sum, and the generating elements interpolate unitarily as follows
\begin{equation}\label{repr_hilbert}
 \Hc=\bigoplus_{n\in\mZ}\Hc_n\,,\qquad
 W(D, c):\Hc_n\mapsto\Hc_{n+n_c}\,.
\end{equation}
With the use of the algebraic relations all matrix elements are expressed in the end in terms of values $\w(W(D, \p^2F))$, as given in \eqref{repr_state}. Therefore, the subalgebra $\Ac_0$ is represented in a regular way, that is one parameter groups $\mR\ni \la\mapsto W(\la D, \la\p^2F)$ are weakly (then also strongly) continuous, so by Stone's theorem (see, e.g.\ \cite{res80}) they are generated by self-adjoint operators:
\begin{equation}\label{repr_FW}
 W(\la D,\la\p^2F)=\exp\big[i\la\Phi(D,\p^2F)\big]\,.
\end{equation}
The algebraic relations of the algebra imply the commutation relations for the generators, satisfied on a suitable domain (which we do not need to specify explicitly for our purposes):
\begin{equation}\label{repr_comPh}
 [\Phi(D_1,\p^2 F_1),\Phi(D_2,\p^2 F_2)]
 =-i\sigma(D_1,\p^2 F_1; D_2, \p^2 F_2)\id\,.
\end{equation}

When restricted to $\Hc_0$, $\pi(\Ac_0)$ is a Fock representation. Namely, let $\Hc_0$ be the Fock space based on the `one excitation' space $\K$ defined above, and identify $\W$ as the Fock-vacuum in that space. For $[G]\in\K$ denote by $d([G])$ and $d^*([G])$ the annihilation and creation operators in that space,
\begin{equation}
 d([G])\W=0\,,\qquad [d([G_1]),d^*([G_2])]=([G_1],[G_2])_\K\,.
\end{equation}
Then
\begin{equation}\label{repr_FF}
 \Phi(D,\p^2F)|_{\Hc_0}
 =\tfrac{1}{\sqrt{2}}\Big\{d([j(D,F)])+d^*([j(D,F)])\Big\}\,,
\end{equation}
with $j(D,F)$ as defined in \eqref{repr_j} (which agrees with \eqref{repr_comPh}). In particular, for $\la\in\mR$ we have $W(\la,0)|_{\Hc_0}=\id$, and by commutation relations we then find
\begin{equation}
 W(\la,0)|_{\Hc_n}=e^{i\la ne}\id\,,\quad \text{so}\quad  W(\la,0)=\exp[i\la Q]\,,
\end{equation}
where $Q$ is a self-adjoint operator with the eigenvalues $ne$ and the corresponding eigenspaces $\Hc_n$ (interpreted as the charge operator).

The representation thus obtained is irreducible. Indeed, let a bounded operator $B$ commute with all $W(D,c)$. Then commutation with $W(\la,0)$ implies that $B\Hc_n\subseteq\Hc_n$. Next, representation of $\Ac_0$ on the Fock space $\Hc_0$ is irreducible, so $B|_{\Hc_0}=\alpha\id$. Finally, $B$ commutes with $W(0,nc_v)$, so $B=\alpha\id$ on the whole space $\Hc$.

The representation is also Lorentz-covariant:
\begin{equation}\label{alg_cov}
 \alpha_\Lambda(W(D,c))=W(T_\Lambda D,T_\Lambda c)=U(\Lambda)W(D,c)U(\Lambda)^*\,,
\end{equation}
with the representation of $U(\Lambda)$ given by
\begin{equation}\label{repr_U}
 U(\Lambda)W(D, c)\W=W(T_\Lambda D, T_\Lambda c)\W\,.
\end{equation}
Each of the spaces $\Hc_n$ is invariant with respect to this representation, so in correspondence to the decomposition \eqref{repr_hilbert}, we have
\begin{equation}\label{repr_Udec}
 U(\Lambda)=\bigoplus_{n\in\mZ} U_n(\Lambda)\,.
\end{equation}

\section{Correspondence with Staruszkiewicz's axioms}\label{corresp}

Here we comment on the correspondence of our Weyl algebra model with Staruszkiewicz's axioms, which was established in Ref.\ \cite{her05}.

We briefly summarize Staruszkiewicz's construction in our language. His classical `phase' field $S(x)$ is the general (save for some regularity demands) solution of the wave equation, homogeneous of degree $0$. Such a field may be represented by\footnote{Here we make use of the representation of the wave equation solutions given in \eqref{app_Z} in Appendix \ref{app_hom}. Expressions $\sgn(x\cdot l)$, $\log|x\cdot l|$, and also $(x\cdot l\pm i0)^{-n}$ to appear below, make sense as distributions in $x$.} (see Eqs.\ (40,41) in \cite{her05})
\begin{equation}\label{corresp_S}
\begin{aligned}
 S(x)&=-\frac{e}{4\pi}\int c(l)\sgn(x\cdot l)d^2 l
 -\frac{e}{4\pi}
 \int\p^2 D(l)\log\Big(\frac{|x\cdot l|}{v\cdot l}\Big)d^2l+S_v\,,\\
 S_v&=\frac{e}{4\pi}\int\frac{D(l)}{(v\cdot l)^2}d^2l\,,
\end{aligned}
\end{equation}
where $c(l)$ and $D(l)$ are smooth functions on the future light cone, homogeneous of degree $-2$ and $0$, respectively; the field $S(x)$ does not depend on the choice of $v\in H$. The constant $S_v$ is interpreted as a phase variable in the reference system with time axis along $v$.

The phase field $S(x)$ gives rise to a Maxwell field, defined in regions outside the cone $x^2=0$, homogeneous of degree $-2$, by
\begin{align}
 F_{ab}(x)&=\nabla_b \Big(\frac{x_aS(x)}{e x^2}\Big)-\nabla_a \Big(\frac{x_bS(x)}{e x^2}\Big)
 =\frac{1}{ex^2}\big[x_a\nabla_bS(x)-x_b\nabla_aS(x)\big]\\[3ex]
 &=\frac{1}{8\pi x^2}\int
 \frac{l_ax_b-l_bx_a}{x\cdot l-i0}
 \Big[\p^2 D(l)-i\tfrac{2}{\pi}c(l)\Big]\,d^2l+\mathrm{c.c.}\,.\label{corresp_K}
\end{align}
In general, this formula is singular on the cone. However, for $c=\p^2F$ there exists a distinguished prescription to regularize the formula and extend it to the whole Minkowski space:  for $x^2\neq0$ integrate $\p^2$ in \eqref{corresp_K} by parts (as in \eqref{app_scalar}), and only then divide by $x^2$. Using the identity
\begin{equation}
 \p^2\,\frac{l_ax_b-l_bx_a}{x\cdot l-i0}
 =-x^2 L_{ab}\frac{1}{(x\cdot l-i0)^2}\,,
\end{equation}
and again integrating $L_{ab}$ by parts, one obtains
\begin{equation}\label{corresp_Ff}
 F_{ab}(x)
 =\frac{1}{8\pi}\int\frac{L_{ab}\big[D(l)-i\tfrac{2}{\pi}F(l)\big]}
 {(x\cdot l-i0)^2}\,d^2l+\mathrm{c.c.}\,,
\end{equation}
which is the general free asymptotic field on Minkowski space, of electric type, cf.\ Eq.\ (18) in \cite{her05}. Note that the term written explicitly in \eqref{corresp_Ff} is a boundary value of an analytic function of $x+iy$, where $y$ is inside the past light cone. Therefore, this term is the positive frequency part of the Fourier representation, while its conjugation is the negative frequency part.

For $\int c(l)d^2l\neq0$ the above procedure for extending \eqref{corresp_K} does not work as it stands. Still, there are (various) ways to regularize $F_{ab}(x)$ also in this case, but this is not what concerns Staruszkiewicz. He restricts attention to the region $x^2<0$ and observes that the field $F_{ab}(x)$ is charged there, with the charge  \begin{equation}\label{corresp_clQ}
 Q=\frac{1}{4\pi}\int c(l)d^2l\,,
\end{equation}
obtained by integrating electric field over a sphere.

Next, the phase field $S(x)$ is quantized heuristically by the substitution  $c\mapsto \hat{c}$, $D\mapsto \hat{D}$, with the condition\footnote{$\hat{D}$ and $\hat{c}$ are now operators, and $D$ and $c$ are test functions.}
\[
 \Big[\int\hat{c}(l)D(l)d^2l,\int\hat{D}(l')c(l')d^2l'\Big]
 =i4\pi\int c(l)D(l)d^2l\,,
\]
Eq.\ (44) in \cite{her05}, which reproduces Staruszkiewicz's commutator, Eq.\ (45) in~\cite{her05}. If one now postulates, still on heuristic level,
\begin{equation}\label{corresp_W}
 W(D,c)=\exp\big[i\sigma(D,c;\hat{D},\hat{c})\big]\,,
\end{equation}
with the condition \eqref{alg_nc} to ensure that the quantized $S_v$ in \eqref{corresp_S} is a phase operator, then these elements satisfy our algebra, which we now regard as the precise formulation of quantization.  For further details we refer the reader to the discussion encompassing Eqs.\ (46-51) in \cite{her05}, where the elements $W(D)$ and $R(c)$ are, in present notation, given by $W(D)=W(D,0)$ and $R(c)=W(0,c)$. This completes the correspondence on the level of algebra.

Now one has to implement Staruszkiewicz's conditions on the state on the algebra, which are formulated in Section 4 of Ref.\ \cite{sta89}. First of all, the state should be Lorentz-invariant, which is satisfied in our construction irrespective of the value $\kappa>0$. Next, our element $W(0,c_v)$ is easily identified with Staruszkiewicz's $\exp[-iS_v]$, and the vector $W(0,c_v)\W$ with his $|v\rangle=\exp[-iS_v]|0\rangle$. The way, in which Staruszkiewicz formulates the remaining conditions, implies that elements $W(D,\p^2F)$ are represented in a regular way, which again is satisfied in all our representations; in Staruszkiewicz's notation $\Phi(D,\p^2F)$ are combinations of the operators $c_{lm}$, $c^{+}_{lm}$ and the charge operator. In particular,
\begin{equation}\label{corresp_qQ}
 \Phi(1,0)=\frac{1}{4\pi}\int \hat{c}(l)d^2l=\hat{Q}
\end{equation}
is the charge operator; compare this formula with the classical charge \eqref{corresp_clQ}. The last, but crucial for the choice of the parameter $\kappa$, of Staruszkiewicz's conditions is that $\W$ is annihilated by charge \eqref{corresp_qQ} and by positive frequency part of fields \eqref{corresp_Ff}. Together, this amounts to the condition
\begin{equation}\label{corresp_Star}
 \big[\p^2\hat{D}(l)-i\tfrac{2}{\pi}\hat{c}(l)\big]\W=0\,.
\end{equation}
This has to be compared with our annihilation operator. Namely, comparing \eqref{corresp_W} with \eqref{repr_FW}, and using \eqref{repr_j}, one obtains for $c=\p^2F$
\begin{equation}
\begin{aligned}
 \Phi(D,\p^2F)
 &=\frac{1}{4\pi}\int\big[D(l)\hat{c}(l)-F(l)\p^2\hat{D}(l)\big]d^2l\\
 &=\frac{i}{8\pi}\int \ov{j(D,F)(l)}
 \big[\kappa^{-\frac{1}{2}}\p^2\hat{D}(l)
 -i\kappa^{\frac{1}{2}}\hat{c}(l)\big]d^2l +\mathrm{h.c.}\,.
\end{aligned}
\end{equation}
Therefore, in the zero charge sector we have by \eqref{repr_FF}:
\begin{equation}
 d([G])=\frac{i}{8\pi}\Big(\frac{2}{\kappa}\Big)^\frac{1}{2}\int \ov{G(l)}
 \big[\p^2\hat{D}(l)-i\kappa\hat{c}(l)
 \big]d^2l\,.
\end{equation}
Agreement with the condition \eqref{corresp_Star} is achieved for $\kappa=2/\pi$, as anticipated in Eq.\ \eqref{repr_kappa}.

\section{Structure of representations $U_n(\Lambda)$}\label{subspaces}

General theory of unitary representations of the universal covering group of the group $\Lor$ was developed by Gelfand et al.\ \cite{ggv66}, with the result of full classification of irreducible representations, grouped in two series, the so-called main- and supplementary series. Moreover, each unitary representation decomposes into irreducibles. Therefore, it is an interesting problem what is a more detailed structure of subrepresentations $U_n(\Lambda)$. Here we gather, with some new additions, known information on that question.

We start by introducing generators of representations \eqref{alg_T} and \eqref{repr_U}. For an infinitesimal Lorentz transformation $\Lambda^a{}_b\approx\delta^a_b+\w^a{}_b$, where $a$, $b$ are spacetime indices, define $L_{ab}$ and $M_{ab}$ by
\begin{equation}\label{subspaces_TU}
 T_\Lambda\approx \id+\tfrac{1}{2}\w^{ab}L_{ab}\,,\qquad
 U(\Lambda)\approx \id-\tfrac{i}{2}\w^{ab}M_{ab}\,,
\end{equation}
which in the case of $T_\Lambda$ acting on scalar functions gives
\[
 L_{ab}=l_a\p_b-l_b\p_a\,,\qquad \p_a\equiv \frac{\p}{\p l^a}\,,
\]
and one easily calculates the Casimir operators of this representation:
\begin{equation}\label{subspaces_casT}
 -\tfrac{1}{2}L^{ab}L_{ab}=(l\cdot\p+2)l\cdot\p\,,\qquad
 \tfrac{1}{2}{}^*\!L^{ab}L_{ab}=0\,,
\end{equation}
where the left star-superscript denotes the dual antisymmetric tensor.

Consider the representation $U_0(\Lambda)$. Its carrier space $\Hc_0$ is the symmetric Fock space based on the one-excitation Hilbert space $\K$. Definition \eqref{repr_U} implies that $U_0(\Lambda)|_{\K}={T_\Lambda}|_{\K}$. But the space $\K$ consists of functions homogeneous of degree $0$, so both Casimir operators \eqref{subspaces_casT} vanish for this restriction. Therefore (see, e.g.,\ formulae (4.3.27,28) in \cite{ohn88}),
\begin{equation}\label{subspaces_eq}
 U_0(\Lambda)|_{\K}\simeq\mathfrak{S}_{2,0}(\Lambda)\,,
\end{equation}
where $\mathfrak{S}_{m, \sigma}$, $m\in\mZ$, $\sigma\in\mR$, is the main series of irreducible, unitary representations (in the notation of \cite{nai64}), and the relation is to be understood as unitary equivalence; in Appendix \ref{app_eq} we give the explicit transformation to the standard form. The tensor powers of this representation, which act in $N$-excitation spaces in $\Hc_0$, decompose into main series with $m\in 2\mZ$ (see the first in the series of three papers by Naimark, Ref.~\cite{nai59}, in which he decomposes tensor products of irreducible representations of $\Lor$).

Structure of representations $U_n(\Lambda)$ for $n\neq0$ is more involved, there is no single building block for them as in the $n=0$ case.
Consider the second Casimir operator
\begin{equation}\label{subspaces_Cas2}
C_2=\tfrac{1}{2}{}^*\!M^{ab}M_{ab}
\end{equation}
of the representation $U(\Lambda)$. By a general theorem for self-adjoint operators we have the orthogonal decomposition
\begin{equation}\label{subspaces_HH}
 \Hc_n=\Hc_n^0\oplus\Hc_n^\times\,,\quad \Hc_n^0=\Ker C_2\cap\Hc_n\,,\quad
 \Hc_n^{\times}=\ov{C_2\Hc_n}\,.
\end{equation}
In Appendix \ref{app_proof} we show the following result. If $n_c=n$, then
\begin{equation}
\begin{aligned}\label{subspaces_c2}
 &C_2W(D,c)\W=0\quad \iff\quad \exists\ v\in H\,,\ x^2<0\,,\ x\cdot v=0\,,\ \la\in\mR:\\[2ex]
 &D=-ne\kappa\arctan\Big[\frac{x\cdot l}{v\cdot l}\Big]+\la,\quad
 c=ne(1-x^2)
 \frac{(v\cdot l)^2-(x\cdot l)^2}{[(v\cdot l)^2+(x\cdot l)^2]^2}\,;
\end{aligned}
\end{equation}
moreover, for $W(D,c)\W$ of this form one has
\begin{equation}\label{subspaces_vz}
 e^{abcd}(v_b+ix_b)M_{cd}W(D,c)\W=0\,.
\end{equation}
In the special case $x=0$, vectors defined by \eqref{subspaces_c2} are of the form $W(0,nc_v)\W$, with $c_v$ as defined in \eqref{alg_cv}.  Relation \eqref{subspaces_vz} is then the statement of their spherical symmetry in the frame with time axis along $v$; cf.\ \cite{sta89}. The closure of the linear span of these vectors forms a Hilbert subspace of $\Hc_n^0$, which we denote $\Hc_n^H$. This leads to further decompositions
\begin{equation}
 \Hc_n^0=\Hc_n^H\oplus\Hc_n^{00}\,,\quad
 \Hc_n=\Hc_n^H\oplus\Hc_n^{00}\oplus\Hc_n^{\times}\,,\qquad (n\neq0)\,,
\end{equation}
where the space $\Hc_n^{00}$ is defined by the first relation. By the above equivalence subspaces $\Hc_n^\times$ are nontrivial. Whether $\Hc_n^{00}$ are  nontrivial is, to our knowledge, an open question. In particular, an interesting problem is whether vectors defined by \eqref{subspaces_c2} with $x\neq0$ are in $\Hc_n^H$.

All subspaces appearing above are invariant with respect to $U_n(\Lambda)$: this is obvious for $\Hc_n^0$ and $\Hc_n^{\times}$, while for $\Hc_n^H$ and $\Hc_n^{00}$ it follows from the obvious relation $T_\Lambda c_v(l)=c_v(\Lambda^{-1}l)=c_{\Lambda v}(l)$. Therefore, this representation has the corresponding decomposition, with subrepresentations denoted with the same superscripts as their carrier spaces.

The representations $U_n^{\times}(\Lambda)$ decompose into irreducible representations with nonzero eigenvalues of $C_2$, that is the main series representations $\mathfrak{S}_{m,\sigma}$ with $m\neq0$, but the details of the decomposition are not known. \emph{A priori}, both series may appear in the decomposition of $U_n^0(\Lambda)$, i.e.\ of $U_n^H(\Lambda)$ and $U_n^{00}(\Lambda)$. Explicit construction of the decomposition of $U_n^H(\Lambda)$, briefly characterized below, has been obtained by Staruszkiewicz \cite{sta92}. On the other hand, to our best knowledge, nothing constructive is known on $U_n^{00}(\Lambda)$.

To achieve the desired decomposition of $U_n^H(\Lambda)$ Staruszkiewicz \cite{sta92} uses the fact that the vectors $W(0,nc_v)\W$, $v\in H$,  form a Lorentz-invariant total set in $\Hc_n^H$, and he diagonalizes the quadratic form kernel on $H\times H$
\begin{equation}\label{subspaces_kernel}
 \big(W(0,nc_v)\W,W(0,nc_u)\W\big)\,.
\end{equation}
The explicit value of this kernel may be easily calculated. Namely, let us denote \begin{equation}\label{subspaces_cc}
 F_{v,u}(l)=e\log\Big[\frac{v\cdot l}{u\cdot l}\Big]\,,\quad
 \text{so that}\quad  \p^2 F_{v,u}(l)=c_u(l)-c_v(l)\,.
\end{equation}
Then the value of the kernel \eqref{subspaces_kernel} follows immediately from the definition of the state \eqref{repr_state} (see Eq.\ (71) in \cite{her05}; Staruszkiewicz's calculation is different)
\begin{equation}\label{subspaces_kerex}
\begin{aligned}
 \big(\W,W(0,n\p^2F_{v,u})\W\big)
 &=\exp\Big[-\frac{\kappa}{4}\|nF_{v,u}\|_\K^2\Big]\\[1ex]
 &=\exp\Big[-\frac{(ne)^2}{\pi}\big(\chi_{v,u}\coth\chi_{v,u}-1\big)\Big]\,,
\end{aligned}
\end{equation}
where $\chi_{v,u}=\arcosh(v\cdot u)$ is the hyperbolic angle between $v$ and $u$, and where in the final expression we substituted~\eqref{repr_kappa}.
Applying the methods of Gelfand et al.\ Staruszkiewicz  succeeded in decomposing $U_n^H$ with the following remarkable result: for $(ne)^2/\pi>1$ it decomposes into a direct integral of main series irreducible representations $\mathfrak{S}_{0,\rho}$ (all of them with zero value of the second Casimir operator), while for $(ne)^2/\pi<1$ there is one single addition of a representation $\mathfrak{D}_\nu$ from the supplementary series, with $\nu=1-ne/\sqrt{\pi}$ (see also \cite{sta04}, \cite{sta09}).

One should mention that Sta\-rusz\-kie\-wicz's diagonalization formula in the form given in \cite{sta92}, which applies to functions $f(u)$ of compact support,  needs an extension to the whole space $\Hc_n^H$, which is necessary for the correctness of the diagonalization in agreement with the general theorem on the direct integral decomposition of the representation $U_n^H(\Lambda)$ (with positive decomposition measure). For this extension, it is necessary and sufficient that the diagonalization measure function $K(\nu;z)$ be nonnegative (notation in this paragraph refers to formulae in \cite{sta92}). This is not immediately obvious from Staruszkiewicz's formula: although we know by the Weyl algebra model that $\langle f|f\rangle\geq0$, the method used for the decomposition of this product is independent of positivity of $K(\nu;z)$, and at the same time the formula does not test its positivity locally, as it relies strongly on analyticity of functions $\check{f}(k;\nu)$, and analyticity of the product in $e^2$. However, the final step assuring positivity of $K(\nu;z)$ is not difficult to be made in the Weyl algebra model. Namely, consider the matrix element $\langle f|\exp[i\la C_1]|f\rangle$, where $C_1$ is the first Casimir operator of the representation $U_n^H$. For $\la\in\mR$ this is the Fourier transform of a positive measure. At the same time, for $\la\in\mC$ with $\Ip\la>0$, and $f$ of compact support, this element is an analytical function of $e^2$ (which is shown by the use of the $H$-Fourier transform, or indeed some explicit formulae of Ref.\ \cite{sta04}), and one can apply the Staruszkiewicz method for its diagonalization. For $\Ip\la\searrow0$ one obtains the Fourier transform of the integrand of the diagonalization of $\langle f|f\rangle$, which therefore must be nonnegative.

The question of the Staruszkiewicz decomposition of the product $\langle f|f\rangle$ has been taken up recently in Ref.\ \cite{waw22}. The author, apparently not aware of the Weyl algebra model assuring consistency of Staruszkiewicz's theory, devotes most of his article to a redundant,  roundabout proof of positivity of the kernel \eqref{subspaces_kerex} (see also remarks in the next section). However, he also gives a proof of the positivity of $K(\nu;z)$ (using the language of positive spherical functions on the Lorentz group and the generalized Bochner theorem), which is probably the first  to have been published explicitly.  The simple proof sketched in the previous paragraph is not related to this advanced method.

We end this section with the observation that the representation $U_n(\Lambda)$ may be unitarily transformed to the space $\Hc_0$. For a fixed  $t\in H$ the element $W(0,nc_t)$ maps $\Hc_0$ unitarily onto $\Hc_n$. Therefore, the formula
\begin{equation}
 U^t_n(\Lambda)=W(0,nc_t)^*U_n(\Lambda)W(0,nc_t)
\end{equation}
defines a unitarily equivalent representation, acting in the Fock space $\Hc_0$. Its generators, by \eqref{app_MW} in Appendix \ref{app_proof}, are
\begin{equation}
 W(0,nc_t)^*M_{ab}W(0,nc_t)=M_{ab}-n\Phi(0,L_{ab}c_t)\,.
\end{equation}
Using this for the definitions \eqref{subspaces_HH} we find what follows. Under the unitary mapping $W(0,nc_t)$ the subspace $\Hc_n^{\times}$ is the image of the closure of the range of the operator
\begin{equation}
 \big[{}^*\!M^{ab}-n\Phi(0,{}^*\!L^{ab}c_t)\big]
 \big[M_{ab}-n\Phi(0,L_{ab}c_t)\big]
\end{equation}
acting in $\Hc_0$, while $\Hc_n^0$ is the image of its kernel.

\section{Comments on references \cite{waw18} and \cite{waw22}}\label{comments}

As shown by the preceding discussion, the mathematical model of the Sta\-rusz\-kie\-wicz theory, formulated as a specific representation of an appropriately chosen Weyl algebra, is precisely defined and cannot raise any doubts about its mathematical consistency (irrespective of the value of $e$). This formulation, recalled in Sections \ref{alg} and \ref{repr} above, was presented in Section 3 of Ref.\ \cite{her05}, based on an earlier Ref.\ \cite{her94}. Therefore, while Ref.\ \cite{waw22} contains a result mentioned in the last section, which is worth noting, most of its contents consists of, in our view, largely redundant and misleading considerations. The main problem the author formulates, is to show `independently' the positivity of kernel \eqref{subspaces_kerex}. In his words: ``In fact, a proof of positivity of [this - AH] kernel, independent of the Staruszkiewicz theory [given in \cite{sta89} - AH], would give us a proof of (relative) consistency of his theory.'' However, in the light of the Weyl algebra model there is no need for any further consistency check. One should also stress that, without a complete consistent model, the positivity of the kernel \eqref{subspaces_kerex} alone, contrary to what the author claims, would by far be not sufficient to prove the consistency of the whole theory. On the other hand, positivity of the kernel in the initial form \eqref{subspaces_kernel} is a trivial fact in the scheme given above, as the kernel is simply the scalar product of vectors in the Hilbert space $\Hc_n^H$, for any value of $e$ and $n$.

Another recent article by the same author, Ref.\ \cite{waw18}, cannot be left without comment either, as it contains a serious mistake leading to false main statements of the article. The~Staruszkiewicz theory not being a subject of my active research, I have only now looked closely at the content of this article. In this article the author claims that, in our notation,
\begin{equation}
 \Hc_n=\Hc_n^H\otimes \Hc_0\,,\quad
 U_n(\Lambda)=U_n^H(\Lambda)\otimes U_0(\Lambda)\,.\qquad
 \text{(false!)}
\end{equation}
This is the content of his lemmas, whose proofs take almost the whole of the article. Basing on the claims of these lemmas, the author formulates unjustified further claims on the decomposition of $U_n(\Lambda)$, going beyond Staruszkiewicz's result mentioned above.

To explain where is the error, we consider vectors of the form
\begin{equation}
 W(D,\p^2F)W(0,nc_v)\W\,,\quad \text{where}\quad
 \sigma(D,\p^2F; 0, nc_v)=0\,,
\end{equation}
which implies that the elements $W(0,nc_v)$ and $W(D,\p^2F)$ commute. The condition on the above vanishing of symplectic mapping $\sigma$ is equivalent to
\begin{equation}\label{rem_com}
 \textstyle\int D(l)c_v(l)\,d^2l=0\,.
\end{equation}
The author claims that there is a unitary equivalence $\Hc_n\mapsto \Hc_n^H\otimes\Hc_0$ under which all such vectors, for all $v$, are transformed as
\begin{equation}
 W(D,\p^2F)W(0,nc_v)\W\mapsto W(0,nc_v)\W\otimes W(D,\p^2F)\W\,.\qquad \text{(false!)}
\end{equation}
If that was true, the scalar products of various vectors of this type, taken after the mapping, should be the same as before the mapping.\footnote{In fact, the claim that this is the case is the main point in author's proof, see formulas (11) and (12) in the article.} Well, there is no reason to believe   this should be true, but let us check this explicitly. Note that these vectors may equally well be written as $W(0,nc_v)W(D,\p^2F)\W$, in which form the condition  \eqref{rem_com} on $D$ may be dropped, as an additive constant in $j(D,F)$ is irrelevant in $\Phi(D,\p^2F)$ \eqref{repr_FF} and $W(D,\p^2F)$ \eqref{repr_FW} on~$\Hc_0$. Let $W(0,nc_{v'})W(D',\p^2F')\W$ be another vector, with $v'\neq v$. Using algebraic relations \eqref{alg_weyl}, definition of the state \eqref{repr_state} and \eqref{repr_gns}, and taking into account relation \eqref{subspaces_cc}, one finds after simple calculation
\begin{align*}
 &\big(W(0,nc_v)W(D,\p^2F)\W, W(0,nc_{v'})W(D',\p^2F')\W\big)\\[1ex]
 &=\big(W(0,nc_v)\W, W(0,nc_{v'})\W\big)\,
 \big(W(D,\p^2F)\W, W(D',\p^2F')\W\big)\times\\[1ex]
 &\hspace{3em}\times\exp\Big\{-i\frac{n}{4\sqrt{2\pi^3}}
 \int(c_{v'}-c_v)\big[\ov{j(D,F)}+j(D',F')]\,d^2l\Big\}\,,
\end{align*}
with $j(D,F)$ defined in \eqref{repr_j}.
The last factor is nontrivial (different from $1$) in general, which contradicts basic relations of \cite{waw18}. For instance, let us replace  $(D,F)$ by $(\la D,\la F)$ and $(D',F')$ by $(\la'D',\la'F')$, $\la,\la'\in\mR$, and calculate $\p^2/\p\la\p\la'$ of the relation for $\la=\la'=0$. The result, with $G=j(D,F)$ and $G'=j(D',F')$, is
\begin{align*}
 &\big(W(0,nc_v)d^*([G])\W, W(0,nc_{v'})d^*([G'])\W\big)
 =\big(W(0,nc_v)\W, W(0,nc_{v'})\W\big)\times\\[1ex]
 &\times\bigg\{\big(d^*([G])\W, d^*([G'])\W\big)
 -\frac{n^2}{16\pi^3} \int(c_{v'}-c_v)\ov{G}\,d^2l
 \int(c_{v'}-c_v)G'\,d^2l  \bigg\};
\end{align*}
the second term in braces is missing in formula (11) in \cite{waw18}. This falsifies not only author's proof. The claim on the tensor product factorization of $U_n$ itself is shown to be wrong.

The equivalence/nonequivalence of representations $U_n$ with differing $n$'s, which was the object of Ref.\ \cite{waw18}, remains unsettled, and depends on the structure of representations $U_n^{00}$ and $U_n^{\times}$.

\renewcommand{\thesubsection}{\arabic{subsection}}

\section{Appendix}\label{app}

\subsection{Homogeneous functions on the light cone}\label{app_hom}

Here we gather a few properties, needed in the main text, of functions $f(l)$ on the future light cone, with definite homogeneity. More information may be found in \cite{her05}.

Let $c(l)$ be homogeneous of degree $-2$. Then the following formula
\begin{equation}
 \int c(l)\,d^2l=\int c(1,\vec{l}\,)\,d\W_t(\vec{l}\,)\,,
\end{equation}
where on the rhs one integrates over the angles in a Minkowski frame with $e_0=t$, defines a Lorentz invariant value, that is for each Lorentz transformation $T_\Lambda$ defined in \eqref{alg_T} one has
\begin{equation}\label{app_inv}
 \int [T_\Lambda c](l)\,d^2l=\int c(l)\,d^2l\,,\quad
 \int L_{ab}c(l)\,d^2l=0\,,
\end{equation}
where $L_{ab}$ are the generators of representation $T_\Lambda$ defined at the beginning of Section \ref{subspaces}.

Denote $\p f(l)=\p f(l)/\p l$, where some extension of $f(l)$ to a neighborhood of the light cone is assumed. For different extensions, $\p f(l)$ restricted to the cone gives different values, which differ by terms proportional to $l$. However, for $D(l)$, $F(l)$ homogeneous of degree $0$ (and extended for the sake of differentiation with the preservation of this property) the expressions $\p^2D(l)$ and $\p D(l)\cdot\p F(l)$ are extension-independent, and the following integral identity holds
\begin{equation}\label{app_scalar}
 \int D\p^2 F\,d^2l=-\int \p D\cdot\p F\,d^2l=\int F\p^2 D\,d^2l\,,
\end{equation}
which may be shown to follow from the second identity in \eqref{app_inv}.
Moreover, one can show that the following equivalence is true:
\begin{equation}\label{app_cF}
 \int c(l)d^2l=0\quad \iff \quad \exists F(l): c=\p^2 F\,,
\end{equation}
where $F$ is homogeneous of degree $0$ and unique up to the addition of a constant.

Finally, we note that for $Z(s,l)$, a function of a real variable $s$ and of $l$~on the future light cone, with homogeneity $Z(\la s,\la l)=\la^{-2}Z(s,l)$, $\la>0$, the integral
\begin{equation}\label{app_Z}
 \int Z(x\cdot l,l)d^2l
\end{equation}
defines a solution of the wave equation. This representation is related to the Fourier representation. For more information see Section 4 in Ref.\ \cite{her05} or Ref.\ \cite{her17}.

\subsection{State $\w$}\label{app_state}

To prove positivity of the linear functional $\w$ defined by \eqref{repr_state}, it suffices to show that $\w(A^*A)\geq0$ for each $\dsp A=\sum_{i=1}^N \xi_iW(D_i,c_i)$, $\xi_i\in\mC$ (see Ref.\ \cite{mstv73}). However, because of the first condition in the definition \eqref{repr_state}, it is sufficient to consider the special case when all $n_{c_i}$ are equal to $n_{c_1}$. Then there exist $F_i$ such that $c_i-c_1=\p^2F_i$, so
$A=W(0,c_1)A'$, where
\begin{equation}
 A'=\sum_{i=1}^N\xi'_i\,W(D_i,\p^2 F_i)\in\Ac_0\,,\quad
 \xi'_i=e^{i\delta_i}\xi_i\,,\quad \delta_i=\frac{1}{8\pi}\int c_1D_i d^2l\,.
\end{equation}
Thus $\w(A^*A)=\w(A^{\prime*}A')$, so the problem is reduced to the subalgebra $\Ac_0$. However, it is easy to see that on that subalgebra we have
\begin{equation}
 \w(W(D,\p^2F))=(\W,\exp[i\Phi(D,\p^2F)]\W)\,,
\end{equation}
with $\Phi(D,\p^2F)$ defined in \eqref{repr_FF}. The rhs defines a Fock state, so the positivity follows. Obviously, $\w(\1)=1$, which closes the proof that $\w$ is a state. Finally, $\|T_\Lambda D\|_\K=\|D\|_\K$ and $\|T_\Lambda F\|_\K=\|F\|_\K$, so the state is Lorentz-invariant.

\subsection{Proof of equivalence (\ref{subspaces_eq})}\label{app_eq}

Here we use the notation of abstract indices \cite{pen84}, in which the spacetime indices are pairs of spinor indices, e.g.\ $a=A\dA$, and $l^a=o^A o^{\dA}$, where $o^A$ is a two-component spinor, and $o^{\dA}$ is its conjugation. For each $G(l)$ a complex,  homogeneous function of degree zero, one has $o^{\dA}\p_{\dA}G(l)=0$ ($\p_{\dA}=\p/\p o^{\dA}$), so we can define $g(o,\bar{o})$ by
\begin{equation}\label{app_Gg}
 \p_{\dA}G(l)=\sqrt{2\pi}\,o_{\dA}g(o,\bar{o})\,,\qquad
 g(\xi o,\bar{\xi}\bar{o})=\bar{\xi}^{-2}g(o,\bar{o})\,,
\end{equation}
where the second relation, with $\xi\in\mC\setminus\{0\}$, shows the homogeneity type of $g$. The Hilbert space of functions $g$ with such homogeneity, equipped with the product $(g_1,g_2)_{2,0}=\int\ov{g_1}g_2\,d^2l$, is the standard carrier space of the representation $\mathfrak{S}_{2,0}(\Lambda)$. We show below that
\begin{equation}\label{app_equivalence}
 ([G_1],[G_2])_\K=(g_1,g_2)_{2,0}\,,
\end{equation}
so the mapping $G\mapsto g$ realizes the equivalence \eqref{subspaces_eq}.

Consider the self-dual part of the generator $L_{ab}$: ${}^+\!L_{ab}=\tfrac{1}{2}(L_{ab}-i{}^*\!L_{ab})$, which in the spinor language has the form\footnote{The reference for these properties is Section 3.4 in \cite{pen84}} ${}^+\!L_{ab}=-\epsilon_{AB}\,o_{(\dA}\p_{\dB)}$.
Using this identity and the definition \eqref{app_Gg}, we find
\begin{equation}\label{app_GGgg}
 \ov{{}^+\!L_{ac}G_1}{}^+\!L_b{}^cG_2=-2\pi\, l_al_b\,\ov{g_1}g_2\,.
\end{equation}
On the other hand, tensorial calculation easily shows that
\begin{equation}
 L_{ac}\ov{G_1}L_b{}^cG_2
 ={}^*\!L_{ac}\ov{G_1}{}^*\!L_b{}^cG_2=l_al_b\,\p \ov{G_1}\cdot\p G_2\,.
\end{equation}
We now choose $t\in H$, contract \eqref{app_GGgg} with $t^at^b$, and divide the resulting equation by $-2\pi(t\cdot l)^2$, which leads to
\begin{align}
 \ov{g_1}g_2&=-\frac{1}{4\pi}\p \ov{G_1}\cdot \p G_2
 -\frac{i}{8\pi(t\cdot l)^2}\big({}^*\!L_{0c}\ov{G_1}L_0{}^c G_2-L_{0c}\ov{G_1}{}^*\!L_0{}^c G_2\big)\\
 &=-\frac{1}{4\pi}\p \ov{G_1}\cdot \p G_2
 -{}^*\!L_{0c}\Big\{\frac{i}{8\pi(t\cdot l)^2}\big(\ov{G_1}L_0{}^c G_2-G_2 L_0{}^c\ov{G_1}\big)\Big\}\,,
\end{align}
where index $0$ indicates contraction with $t$, and in the second step we used easily verifiable identities ${}^*\!L_{0c}t\cdot l=0$,
${}^*\!L_{0c}L_0{}^c G_i=0$. Integrating this identity we arrive at \eqref{app_equivalence}.

\subsection{Proof of equivalence (\ref{subspaces_c2}) and identity (\ref{subspaces_vz})}\label{app_proof}

With the use of relations \eqref{alg_cov}, \eqref{subspaces_TU} and \eqref{alg_weyl}, one verifies that
\begin{align}
 \big[M_{ab},W(D,c)\big]&=W(D,c)\Big\{-\Phi(L_{ab}D,L_{ab}c)
 +m_{ab}(D,c)\Big\}\,,\label{app_MW}\\
 m_{ab}(D,c)
 &\equiv\tfrac{1}{2}\sigma(D,c;L_{ab}D,L_{ab}c)=\frac{1}{4\pi}\int DL_{ab}c\,d^2l\,,\\[1ex]
 \big[M_{ab},\Phi(D,c')\big]&=i\Phi(L_{ab}D,L_{ab}c')\,,\label{app_MPhi}
\end{align}
where the last relation applies in the case $n_{c'}=0$.
Using~these relations, for $C_2$ the second Casimir operator  \eqref{subspaces_Cas2},  one obtains
\begin{equation}\label{app_comCas}
\begin{aligned}
 C_2W(D,c)=\tfrac{1}{2}W(D,c)\Big\{{}^*\!M^{ab}-\Phi({}^*\!L^{ab}D,{}^*\!L^{ab}c)
 +{}^*\!m^{ab}(D,c)\Big\}&\times\\
 \times\Big\{M_{ab}-\Phi(L_{ab}D,L_{ab}c)
 +m_{ab}(D,c)\Big\}&\,.
\end{aligned}
\end{equation}
Applying this relation to the vector $\W$, and taking into account \eqref{app_MPhi} and the second identity in \eqref{subspaces_casT} to commute ${}^*\!M^{ab}$ with $\Phi(L_{ab}D,L_{ab}c)$, one finds
\begin{equation}
\begin{aligned}\label{app_C2WDc}
 C_2W(D,c)\W=\tfrac{1}{2}
 W(D,c)\Big\{\Phi({}^*\!L^{ab}D,{}^*\!L^{ab}c)
 -{}^*\!m^{ab}(D,c)\Big\}&\times\\
 \times\Big\{\Phi(L_{ab}D,L_{ab}c)
 -m_{ab}(D,c)\Big\}&\W\,.
\end{aligned}
\end{equation}
Next, one notes that if $n_c=n$, then $c=nc_u+\p^2F$, with an arbitrarily fixed $u\in H$, and the associated function $F$. It is now easy to show that \[
 L_{ab}c=\p^2(l_af_b-l_bf_a)\,,\quad \text{where}\quad
 f_b=-ne\frac{u_b}{u\cdot l}+\p_bF\,.
\]
Using relation \eqref{repr_FF}, we obtain
\begin{gather}
 \Phi(L_{ab}D,L_{ab}c)|_{\Hc_0}=\tfrac{1}{\sqrt{2}}\Big\{d([h_{ab}])+d^*([h_{ab}])\Big\}\,,\label{app_Phi}\\[-2ex]
 \intertext{where}
 h_{ab}=j(L_{ab}D, l_af_b-l_bf_a)=l_ak_b-l_bk_a\,,\\[1ex]
 k_b=\p_b\, j(D,F)+ine\kappa^{\frac{1}{2}}\frac{u_b}{u\cdot l}\,,\qquad k\cdot l=ine\kappa^\frac{1}{2}\,.\label{app_k}
\end{gather}
We note that $\p_ak_b-\p_bk_a=0$, so $\p^2h_{ab}=l_a\p^2k_b-l_b\p^2k_a$, and in consequence,
\[
 [d([{}^*\!h^{ab}]),d^*([h_{ab}])]=([{}^*\!h^{ab}],[h_{ab}])_\K=0\,.
\]
Therefore, the rhs of \eqref{app_C2WDc} takes the form: $\frac{1}{2}W(D,c)$ applied to
\begin{equation}
 \Big\{\tfrac{1}{2}d^*([{}^*\!h^{ab}])d^*([h_{ab}])\label{app_dd}
 -\sqrt{2}\,{}^*\!m^{ab}d^*([h_{ab}])+{}^*{}\!m^{ab}m_{ab}\Big\}\W\,.
\end{equation}

Suppose that $C_2W(D,c)\W=0$. Vanishing of the first term in \eqref{app_dd} means that the symmetrical function  $G(l_1,l_2)=e^{abcd}h_{ab}(l_1)h_{cd}(l_2)$ is the zero vector in the symmetrized Hilbert space $\K\otimes_s\K$. This is equivalent to the existence of a function $J(l)$ such that $G(l_1,l_2)=J(l_1)+J(l_2)$. However, $G(l,l)=0$, so $J(l)=0$. We arrive at the following consequence:
\begin{equation}\label{app_ll}
 l_1\wedge k(l_1)\wedge l_2\wedge k(l_2)=0\qquad \forall\ l_1, l_2\,.
\end{equation}
We fix arbitrary $l_0$ and $k_0\equiv k(l_0)$, and note that $l_0\wedge \Ip k_0\neq0$ by the second relation in \eqref{app_k}, so also $l_0\wedge k_0\neq0$. For each $l$ such that $l_0\wedge k_0\wedge l\neq0$, we then have \mbox{$k(l)=\alpha(l)k_0+\beta(l)l_0+\gamma(l)l$}. As $k(l)$ is defined up to the addition of a term proportional to $l$, we can replace it by   $\tilde{k}(l)=\alpha(l)k_0+\beta(l)l_0$. Consider now the set of pairs $l_1, l_2$ such that $l_0\wedge \Ip k_0\wedge l_1\wedge l_2\neq0$; such pairs form a dense set in the set of all pairs. For these pairs equation \eqref{app_ll} gives
\begin{equation}
 l_1\wedge l_2\wedge k_0\wedge l_0\big[\alpha(l_1)\beta(l_2)-\alpha(l_2)\beta(l_1)\big]=0\,,
\end{equation}
and the wedge product in front of the bracket does not vanish.
It follows that $(\alpha(l),\beta(l))=\sigma(l)(\gamma, \delta)$, where $(\gamma, \delta)$ is a constant pair, so $\tilde{k}(l)=\sigma(l)\xi$, where $\xi=\gamma k_0+\delta l_0$ is a constant complex vector.  Using the second equality in \eqref{app_k}, we find $ine\kappa^\frac{1}{2}=\sigma(l)\xi\cdot l$, which leads to
\begin{equation}\label{app_kxi}
 \tilde{k}(l)=ine\kappa^\frac{1}{2}\frac{\xi}{\xi\cdot l}\,.
\end{equation}
As the function $\tilde{k}(l)$ has to be smooth by assumption, the modulus  $|\xi\cdot l|$ cannot vanish, which implies that the vectors $\Rp\xi$, $\Ip\xi$ span a timelike $2$-surface or a timelike straight line.  Multiplying $\xi$ by a suitable complex number one can bring this vector (without changing $\tilde{k}$) to the form $\xi=v+ix$, where $v\in H$ and \mbox{$x\cdot v=0$}. If we identify functions $D$ and $F$ by the condition
\begin{equation}\label{app_jDF}
 j(D,F)=ine\kappa^\frac{1}{2}\log\Big[\frac{\xi\cdot l}{v\cdot l}\Big]+\con\,,
\end{equation}
then $\tilde{k}$ given by \eqref{app_kxi} takes the form \eqref{app_k} with $v$ replacing $u$, which proves its admissibility. Extracting $D$ and $F$ from \eqref{app_jDF}, and then calculating \mbox{$c=nc_v+\p^2F$}, one obtains $(D,c)$ as given in the equivalence \eqref{subspaces_c2}. This completes the proof `from left to right'.\pagebreak[2]

For implication `from right to left' we observe that given $D$ and $c$ as in \eqref{subspaces_c2}, one can recover vector $k$ in the form \eqref{app_kxi}, with $\xi=v+ix$, and it then follows easily that the first term in \eqref{app_dd} vanishes. Moreover, as $D$ and $c$ are determined by only two constant vectors $v$ and $x$, the antisymmetric tensor quantity $m_{ab}(D,c)$ is proportional to $v_ax_b-v_bx_a$. It is now easy to see that also the remaining terms in \eqref{app_dd} vanish. This completes the proof of equivalence \eqref{subspaces_c2}.

Finally, it is now evident that for $W(D,c)$ satisfying the conditions of the equivalence, the antisymmetrization with $\xi=v_c+ix_c$ of the rhs of \eqref{app_MW} applied to $\W$ gives zero. The identity  \eqref{subspaces_vz} follows.

\end{document}